\documentclass{article} %

\usepackage{PRIMEarxiv}
\usepackage[utf8]{inputenc} 
\usepackage[T1]{fontenc}    
\usepackage{hyperref}       
\usepackage{url}            
\usepackage{booktabs}       
\usepackage{amsfonts}       
\usepackage{nicefrac}       
\usepackage{microtype}      
\usepackage{lipsum}
\usepackage{fancyhdr}       
\usepackage{graphicx}       
\graphicspath{{media/}}     
\usepackage{amsmath} 
\usepackage{pdfx}
\usepackage{hyperxmp}

\title{
Deep Operator Learning for High-Fidelity Fluid Flow Field Reconstruction from Sparse Sensor Measurements
}

\author{
  Hiep V. Dang \\
  University of Virginia,\\
  Charlottesville, VA 22903, United States \\
  \texttt{zgp2ps@virginia.edu} \\
  \And
  Phong C. H. Nguyen\thanks{Corresponding author} \\
  Phenikaa University \\
  Hanoi 100000, Vietnam\\
  \texttt{phong.nguyenconghong@phenikaa-uni.edu.vn}
}

\begin{document}
\maketitle

\begin{abstract}
Reconstructing high-fidelity fluid flow fields from sparse sensor measurements is vital for many science and engineering applications but remains challenging because of the dimensional disparities between state and observational spaces. Due to such dimensional differences, the measurement operator becomes ill-conditioned and non-invertible, making the reconstruction of flow fields from sensor measurements extremely difficult. Although sparse optimization and machine learning address the above problems to some extent, questions about their generalization and efficiency remain, particularly regarding the discretization dependence of these models. In this context, deep operator learning offers a better solution as this approach models mappings between infinite-dimensional function spaces, enabling superior generalization and discretization-independent reconstruction. We introduce a deep operator learning model that is trained to reconstruct fluid flow fields from sparse sensor measurements. Our deep learning model employs a branch-trunk network architecture to represent the inverse measurement operator that maps sensor observations to the original flow field, a continuous function of both space and time. Our validation has demonstrated that the proposed deep learning method consistently achieves high levels of reconstruction accuracy and robustness, even in scenarios where sensor measurements are inaccurate or missing. Furthermore, the operator learning approach enables the capability to perform zero-shot super-resolution in both spatial and temporal domains, offering a solution for rapid reconstruction of high-fidelity flow fields.
\end{abstract}


\section{Introduction}
\subsection{Background}
\begin{figure*}[tb!]
    \centering
    \includegraphics[width=0.65\textwidth]{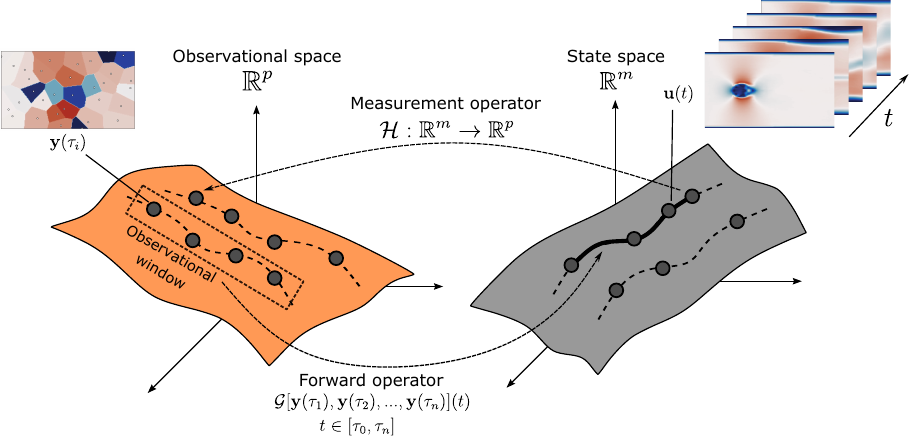}
    \caption{{Problem formulation of FLRONet.} FLRONet utilizes deep operator network to reconstruct the original flow field evolution for a given observation time window $\left[\tau_0, \tau_n \right]$. The reconstruction is continuous in both spatial and temporal domains.}
    \label{fig:1}
\end{figure*}

\label{sec:level1}
The ability to reconstruct the fluid flow field from limited sensor measurements is of paramount importance in numerous scientific and engineering applications. In this problem, the objective is to reconstruct the high-fidelity flow field $\mathbf{u} (t)$ from a sparse set of sensor observations $\mathbf{y}(t)$. Here, both $\mathbf{u} (t)$ and $\mathbf{y}(t)$ are functions of time and return values in $\mathbb{R}^m$ and $\mathbb{R}^p$ (with $m \gg p$), respectively. This task is typically difficult due to the distinctive characteristics of the measurement operator \(\mathcal{H}: \mathbb{R}^m \to \mathbb{R}^p\) that maps a specific state of the flow field $\mathbf{u}$ at time $t = \tau$ to its corresponding observations $\mathbf{y}$. Due to the large discrepancy between the dimension of $\mathbb{R}^m$, \textit{i.e.,} the \textit{state space}, and the dimension of $\mathbb{R}^p$, \textit{i.e.,} the \textit{observational space}, the measurement operator $\mathcal{H}$ is often ill conditioned and thus is non-invertible. Therefore, it is nearly impossible to derive the forward map $\mathcal{G}: \mathbb{R}^p \to \mathbb{R}^m$ as the inverse of $\mathcal{H}$. For this reason, alternative methodologies such as data-driven approaches \cite{Erichson2020}, sparse optimization \cite{Loiseau_Noack_Brunton_2018}, and data assimilation \cite{MONS2016255} are frequently adopted. However, despite significant progress reported in the literature, the reconstruction of fluid flow fields from sparse sensor measurements remains a formidable and unresolved challenge.

The emergence of machine learning has significantly transformed the research in flow field reconstruction from sparse sensor data. Evidences in the literature prove that deep neural surrogates has surpassed the accuracy–cost constraints associated with conventional modal decomposition methods and direct numerical simulation (DNS). Notably, Li \textit{et al.} \cite{li2024deep} demonstrated that a residual CNN trained on direct numerical simulation data, is capable of recovering high-Re cylinder wakes with superior accuracy compared to proper orthogonal decomposition (POD) or dynamic mode decomposition (DMD) while achieving a reduction in inference time by  multiple orders of magnitude. Similarly, Erichson \textit{et al.} \cite{Erichson2020} employed a shallow auto-encoder that realized comparable efficiencies, providing rapid and accurate reconstructions using only eight sensors. Several more advanced approaches that utilize convolutional neural networks have reaffirmed the efficacy of deep learning. For example, Peng \textit{et al.} \cite{PENG2023108539} integrated multiscale U-Net blocks with frequency–residual learning techniques to refine the reconstruction of unsteady aerodynamic fields beyond the capabilities of reduced-order surrogates. Additionally, Liu \textit{et al.} \cite{LIU2021120684} utilized a physics-guided CNN to enhance the accuracy of nanofluid heat-transfer field reconstruction. Experimental validations in industrial contexts substantiate these findings. Fukami \textit{et al.} \cite{Fukami} employed supervised machine learning methods to reconstruct turbulent vortical structures within a pump sump, achieving precise flow estimations utilizing merely a limited number of sensor measurements. More recently, Zhu \textit{et al.} \cite{Zhu2025} incorporated finite-volume constraints within a physics-informed U-Net to effectively capture cylinder-wake dynamics using only a small number of sensors. Furthermore, beyond direct inversion, deep networks serve as mesh-free implicit representations facilitating sparse optimization \cite{luo2024continuous} and operate as dynamic priors within data-assimilation frameworks, where they are continuously refined by incoming measurements \cite{CHENG2024112581}.

Despite these advances, most deep-learning reconstructions continue to encounter challenges in transferring effectively across different meshes or sensor configurations. The primary reason for this is the networks' usual training at a single grid resolution, whereby even minimal alterations in discretization or sensor positioning require expensive retraining and hyper-parameter optimization. Within this context, two principal obstacles emerge. First, the acquisition of large, high-resolution datasets essential for re-training such models—originating either from laboratory experiments or DNS—remains costly, thereby relegating numerous real-world problems to a “small-data” domain that intensifies overfitting and diminishes reliability across conditions. Second, in scenarios where sufficient data are accessible, the training of state-of-the-art networks on refined grids becomes computationally demanding, resulting in prolonged turnaround times and significant energy consumption—limitations that are misaligned with real-time or embedded applications. These constraints inherent to mesh-dependent problems were in fact delineated by Nguyen \textit{et al.} \cite{Nguyen_review} and continue to impede the regular use of deep learning for flow reconstruction within practical engineering contexts.

Operator-learning frameworks address these challenges by learning the maps between infinite-dimensional function spaces, as opposed to fixed-length state vectors. Neural operators such as the Graph Kernel Network and the Fourier Neural Operator (FNO) are now acknowledged as discretization-invariant surrogates for partial differential equations (PDEs) \cite{Li2020MultipoleGN, JMLR:v24:21-1524}. Because the learned kernel operates in the spectral domain, the same weights can be applied to meshes of arbitrary refinement, enabling zero-shot super-resolution—a capability that standard CNN or POD models lack. Recent reviews underscore the value of such mesh independence advantage for scientific computing and engineering applications \cite{Faroughi2024, Moya2023}. Preliminary studies indicate that operator learning models, by virtue of their inherent inductive biases, not only generalize across different discretizations but also achieve superior data efficiency and reduced inference costs compared to traditional DNS surrogates \cite{Li2020MultipoleGN,Wen2021UFNOA,Li2021PhysicsInformedNO}. These characteristics render operator learning particularly advantageous for engineering datasets characterized by non-uniform meshes or sparse sensor graphs, contexts in which conventional modal models or convolutional neural networks necessitate redesign for each new discretization.

Building on FNO’s mesh-invariant promise, Zhao \textit{et al.} \cite{ZHAO2024108619} introduced RecFNO, an FNO-based inverse model that reconstructs transient heat-flow fields from sparse thermocouples and attains up-to-$8\times$ spatial super-resolution without retraining.  Comparable gains were reported by Yao \textit{et al.} \cite{Yao2023}. The authors leveraged an FNO backbone to recover time-varying wing-surface pressure distributions with high accuracy and strong noise immunity. These successes, however, address only the spatial half of the reconstruction problem—each snapshot is treated independently, and temporal coherence is left to chance.  In principle, one might extend FNO to the three dimensions by appending time as an extra spatial axis, but the spectral tensor then scales cubically with the number of steps, quickly exhausting GPU memory and confining predictions to short horizons. Existing FNO reconstructions are therefore discretization-independent in space yet fragile in time, a gap that motivates the nested spatio-temporal operator proposed in this work.

The deep operator network (DeepONet) offers a complementary route to mesh-free reconstruction by learning operators directly from scattered function samples to continuous outputs.  Lu \textit{et al.} \cite{lu2021learning} proved that its branch–trunk architecture generalizes the universal approximation theorem to the operator level, effectively acting as a data-driven Green’s function. Subsequent studies have validated this claim across diverse physics domains—from bubble growth \cite{lin2021operator} and crack predictions \cite{goswami2022physics} to geometric deformation estimation \cite{Liu2025}. In each case, DeepONet delivers accurate, physics-consistent predictions while remaining agnostic to the underlying mesh. This discrete-to-continuous mapping is particularly well-matched to sparse-sensor flow reconstruction, where probe values form the branch inputs and the sought-after space–time field emerges from the trunk evaluation at arbitrary query points.

\subsection{Contributions}
Leveraging recent advances in operator learning, we present FLRONet, a nested spatio-temporal deep operator network meticulously crafted to reconstruct fluid flow fields from sparse sensor measurements and being independent of the discretization across spatial and temporal domains. At its core, FLRONet decomposes the inverse measurement operator into two complementary components: (i) a spatial branch network, implemented via a two-dimensional (2D) FNO, which maps each sparse sensor snapshot to a compact, discretization-invariant latent representation; and (ii) a temporal trunk network, takes these latent representations as input to establish correlations with the designated temporal domain by employing the Green's function formulation, thereby enabling continuous flow-field reconstructions at arbitrary query times.

This formulation yields an operator-learning method distinguished by three critical properties, which is, to the best of our knowledge, have not previously been achieved concurrently in existing methods. First, FLRONet supports \textit{dual-domain zero-shot super-resolution}, providing seamless reconstruction at spatial and temporal resolutions significantly finer than those seen during training. Specifically, as demonstrated in Section \ref{sec:results}, FLRONet trained only on a coarse grid $140\times240$ can reliably predict detailed flow structures at resolutions as fine as $1120\times1920$, or at intermediate time steps smaller than $\Delta t < 10^{-6}\text{s}$, without retraining. Second, the model benefits from inductive biases inherent in FNO and DeepONet architectures, enabling robustness against sensor sparsity and measurement noise. By incorporating a Voronoi embedding to handle incomplete sensor data and employing Fourier-based spectral filtering to mitigate high-frequency disturbances, FLRONet achieves stable and accurate predictions even when confronted with severe sensor failures (up to 50$\%$) or substantial noise levels (up to 20$\%$ Gaussian noise). Lastly, FLRONet provides substantial improvements in computational speed. It achieves real-time inference at roughly $16\text{ms/frame}$ on an A100 GPU, outperforming competing 3-D FNO models in both reconstruction accuracy and speed. Collectively, these attributes position FLRONet as an effective, mesh-agnostic, and computationally efficient digital-twin surrogate, significantly extending the discretization-invariant promise of neural operators \cite{DBLP:journals/corr/abs-2010-08895, lu2021learning} into the challenging realm of inverse problems associated with sparse-sensor flow reconstruction.

\section{Problem formulation}

Our problem formulation is illustrated in Fig. \ref{fig:1}. Given a fluid system, the state of the flow field is represented as a time-dependent function $\mathbf{u}(t)$ that returns values in $\mathbb{R}^m$. Measurements $\mathbf{y} \in \mathbb{R}^p$ are acquired at specified discrete-time instances $\tau_i \in \{\tau_0, \tau_1, ..., \tau_n \}$ and taken at $p$ locations within the fluid domain. This process is represented by a measurement operator $\mathcal{H}: \mathbb{R}^m \to \mathbb{R}^p$:

\begin{equation}
\label{eqn:measurement_op}
    \mathbf{y} = \mathcal{H} [\mathbf{u}].
\end{equation}

In this formulation, $\mathbb{R}^m$ is referred to as the \textit{state space}, while $\mathbb{R}^p$ denotes the \textit{observational space}. Our goal here is to reconstruct the original flow field $\mathbf{u}$ from sparse sensor measurements $\mathbf{y}$ at a specified time $t \in [\tau_0, \tau_n]$. This corresponds to a temporal \textit{interpolation} problem, where the model infers flow fields at arbitrary continuous time points strictly within the observation window. Note that the process of temporal \textit{extrapolation} after the last observation ($t > \tau_n$) is beyond the scope of this study and will be investigated in our future work. 

The reconstruction task described above is fundamentally equivalent to determining the forward operator $\mathcal{G}: \mathbb{R}^p \to \mathbb{R}^m$, which theoretically is the inverse of the measurement operator $\mathcal{H}: \mathbb{R}^m \to \mathbb{R}^p$:

\begin{equation}
\label{eqn:forward_op}
    \mathbf{u} = \mathcal{G}[\mathbf{y}] \approx \mathcal{H}^{-1}[\mathbf{y}] .
\end{equation}

In contrast to available methodologies that focus on modeling $\mathcal{G}$ as a mapping between sensor observations and the state field at a single discrete time point, our approach seeks to incorporate the dynamic properties of fluid flow into the reconstruction process. This integration imposes an additional constraint on the learning process, ensuring the integrity of the reconstructed field. Moreover, we also want our reconstructed fields to be independent of the selected discretization settings. Therefore, our objective is to reconstruct the state field $\mathbf{u}$ given a set of sensor observations at discrete time points $\left[ \mathbf{y}(\tau_0), \mathbf{y}(\tau_1), ..., \mathbf{y}(\tau_n) \right]$ for any given time $t \in \left[\tau_0, \tau_n \right]$:

\begin{equation}
\label{eqn:deeponet}
\mathbf{u}(t) = \mathcal{G}[\mathbf{y}(\tau_1), \mathbf{y}(\tau_2), ..., \mathbf{y}(\tau_n)](t) .
\end{equation}

with $\mathbf{u}(t)$ is discretization independent in both space and time. With the above objective and inspired by the deep operator theory of Lu \textit{ et al.} \cite{lu2021learning}, we approximate the operator $\mathcal{G}$ by a deep operator network for which we named FLRONet.

\section{Methodology}
\label{sec:method}

\begin{figure*}[thb!]
    \centering
    \includegraphics[width=.8\textwidth]{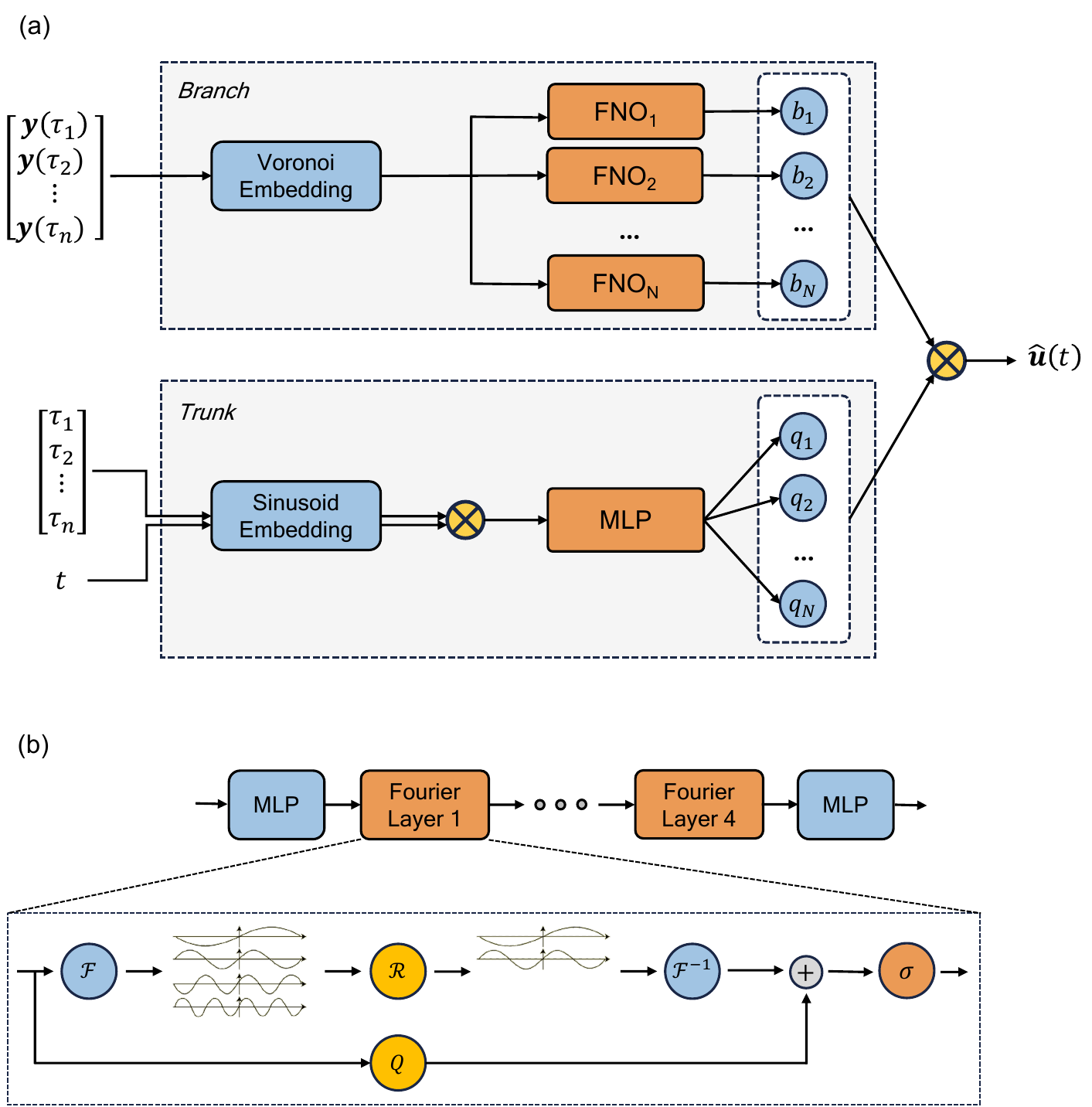}
    \caption{{(a) Architectural design of FLRONet.} FLRONet architecture employs a branch-trunk framework. The \textbf{branch network} encodes the spatial relationship among sensors in space by processing  $[\mathbf{y}(\tau_1), \mathbf{y}(\tau_2), ..., \mathbf{y}(\tau_n)]$ through an Voronoi embedding layer $\Theta$. The resulting embeddings are then passed to $N$ independent two-dimensional FNO blocks, generating $N$ output fields $b$. The $\textbf{trunk network}$ models the temporal correlation between the target time $t$ and the observation time instances $[\tau_1, \tau_2, ..., \tau_n]$. It applies a sinusoid embedding layer $\Phi$ to embed $t$ and $\tau$ independently, combining them via a dot product, and finally passes the result to an MLP to produce $N$ output values $q$. A dot product combines $b$ and $q$ to produce the final construction $\mathbf{\hat{u}}(t)$.
    {(b) Architecture of each FNO block.} We adopt the standard configuration with four Fourier layers. Each layer applies a discrete Fourier transform $\mathcal{F}$ to map the physical input into the frequency domain, a complex-valued linear transformation $\mathcal{R}$ performs frequency processing. Afterward, only the lowest-frequency modes are retained before being transformed back to physical domain via the inverse Fourier transform $\mathcal{F}^{-1}$. A direct linear transformation $\mathcal{Q}$ is also added to the output before going to the non-linear activation $\sigma$.
    }
    \label{fig:2}
\end{figure*}

\subsection{Architecture design}
\label{subsec:design}
Fig. \ref{fig:2} illustrates the architecture of FLRONet, which comprises two separate networks: the \textit{trunk} and the \textit{branch}. The branch network encodes the spatial relationship among the sensors in space by processing $[\mathbf{y}(\tau_1), \mathbf{y}(\tau_2), ..., \mathbf{y}(\tau_n)]$ through a Voronoi embedding layer $\Theta$. The resulting embeddings are then passed to $N$ independent two-dimensional FNO blocks, generating $N$ output fields $b$. The trunk network models the temporal correlation between the target time $t$ and the observation time instances $[\tau_1, \tau_2, ..., \tau_n]$. It applies a sinusoid embedding layer $\Phi$ to embed $t$ and $\tau$ independently, combining them via a dot product, and finally passes the result to an MLP to produce $N$ output values $q$. A dot product combines $b$ and $q$ to produce the final construction $\mathbf{\hat{u}}(t)$.

\subsection{Operator networks for spatiotemporal discretization independent}
\subsubsection{DeepONet with FNO branch net}
The primary contribution of our work lies in the ability to reconstruct the evolution of fluid flow fields from sparse sensor measurements that is independent of discretization in both the spatial and temporal domains. This capability is achieved through a DeepONet-like architecture. In particular, the forward operator $\mathcal{G}$ can be modeled as follows:

\begin{equation}
\label{eq:model}
\mathcal{G}(\mathbf{y}(\tau))(t) = \mathcal{B}(\mathbf{y}(\tau)) \otimes \mathcal{T}(t; \tau).
\end{equation}

In Eq. \ref{eq:model}, the branch network $\mathcal{B}$ processes sensor measurements $\mathbf{y}$ at discrete time instances $\tau$, while the trunk network $\mathcal{T}$ receives the target reconstruction time $t$ and $\tau$ as input. This formulation of $\mathcal{T}$ aims to elucidate the relative proximity between the reconstruction time $t$ and the observation time $\tau$. The output of the branch and trunk networks is combined by a fusion operation.

In our work, $\mathcal{T}$ is modeled by an MLP with three fully connected layers, preceded by the sinusoidal embedding $\Phi$ to improve the representation of relative temporal positions. The sinusoidal embeddings of $\tau$ and $t$ are combined using a simple dot product between two vector spaces. The dot product increases in response to temporal proximity. This design allows the trunk network to effectively capture the temporal relationship between the sensor input and the reconstruction target.

The branch network $\mathcal{B}$ is designed on top of the unstacked version of DeepONet \cite{lu2021learning}. Although the branch network can be implemented using various architectures, i.e., U-Net \cite{Ronneberger2015} or fully connected networks \cite{lu2021learning}, in our work, we use $N$ independent blocks of Fourier Neural Operators to attain spatial resolution independence.

With this design, Eq. \ref{eq:model} becomes:

\begin{equation}
    \label{eq:flronet}
    \mathcal{G}(\mathbf{y}(\tau))(t) = 
    \displaystyle\sum_{i=1}^{N} \langle\text{FNO}_i\left[\Theta(\mathbf{y})\right], \text{MLP}\left[\langle\Phi(t), \Phi(\tau)\rangle\right]\rangle 
\end{equation}

where $\langle\cdot, \cdot\rangle$ refers to the dot product between two vector spaces.

\subsubsection{FNO branch net}

The input to each FNO branch is $\Theta(\mathbf{y})$, the Voronoi embedding of $y$. As will be detailed in Section \ref{subsec:voronoi}, the Voronoi layer $\Theta$ computes the embedding at every spatial coordinate $x$ using nearest-neighbor interpolation on sparse sensor values. Therefore, $v = \Theta(\mathbf{y})$ is a function of $x$ and a Fourier Neural Operator can be naturally applied on $v$ to model the fluid flow over space. Each Fourier layer in Fig. \ref{fig:2} is now delineated as:

\begin{equation}
\label{eq:fno}
\text{FNOLayer}(v(x)) = \sigma\left(\mathcal{F}^{-1} (\mathcal{F}(\kappa) \cdot \mathcal{F}(v))(x) + \mathcal{Q}(v(x))\right) ,
\end{equation}

where $\sigma$ is a nonlinear activation, $\mathcal{Q}$ is a linear transform, $\kappa$ is the kernel function, $\mathcal{F}$ and $\mathcal{F}^{-1}$ are the Fourier transform and its inverse. By definition, in the two-dimensional spatial domain $\Omega$, the Fourier transform and its inverse are given as:

\begin{figure}[h!]
\centering
\begin{minipage}{0.47\textwidth}
\begin{equation}
\label{eq:continous_fourier}
\hat{v}(k) = \mathcal{F}v = \int_{\Omega} f(x) e^{- i 2 \pi \langle k , x \rangle} \, dx ,
\end{equation}
\end{minipage}\hfill
\begin{minipage}{0.47\textwidth}
\begin{equation}
\label{eq:inv_continous_fourier}
v(x) = \mathcal{F}^{-1}f = \int_{\mathbb{R}^2} \hat{v}(k) e^{i 2 \pi \langle k , x  \rangle} \, dk.
\end{equation}
\end{minipage}
\end{figure}

Considering that we are restricted to evaluating the value $\mathbf{\hat{u}}(t)$ on a finite mesh, Eq. \ref{eq:continous_fourier} and \ref{eq:inv_continous_fourier} can be discretized using the two-dimensional discrete Fourier transform (DFT) and the inverse discrete Fourier transform (IDFT). Assume that the spatial domain $\Omega$ is discretized into a uniform grid, with $N_1$ and $N_2$ denoting the number of grid points along two axes. The two-dimensional DFT and IDFT are given by:
\begin{equation}
\begin{split}
   \hat{v}[k_1, k_2] = \text{DFT}(v)= \Delta x_1 \Delta x_2 \sum_{n_1=0}^{N_1-1} \sum_{n_2=0}^{N_2-1} v[n_1 \Delta x_1, n_2 \Delta x_2]\,e^{- i 2 \pi \left( \frac{k_1 n_1}{N_1} + \frac{k_2 n_2}{N_2} \right)}, 
\end{split}
\label{eq:discrete_fourier}
\end{equation}
\begin{equation}
\begin{split}
v[x_1, x_2] = \text{IDFT}(v) = \frac{1}{N_1 N_2 \Delta x_1 \Delta x_2} \sum_{k_1=0}^{N_1-1} \sum_{k_2=0}^{N_2-1} \hat{v}[k_1, k_2]\,e^{i 2 \pi \left( \frac{k_1 n_1}{N_1} + \frac{k_2 n_2}{N_2} \right)} .
\end{split}
\label{eq:inv_discrete_fourier}
\end{equation}
A key principle of FNO is to retain only low-frequency modes for further processing while truncating the rest. This principle is inspired by Parseval's Theorem, which implies that low-frequency modes of a function capture most of its information while high-frequency modes typically contain redundant details or noise. For our $140 \times 240$ spatial domain, we retain up to $24$ modes along the vertical axis ($k_1 = 24$) and $48$ modes along the horizontal axis ($k_2 = 48$). 

Lastly, we adhere to the standard parameterization of $\mathcal{F}(\kappa)$: Given $d$ as the embedding dimension, $\mathcal{F}(\kappa)$ is defined by a complex-valued linear transform $\mathcal{R}$, with parameters $\phi_{\mathcal{R}} \in \mathbb{C}^{k_1 \times k_2 \times d \times d}$. In this study, we choose $d=64$.

\subsection{Voronoi tessellation for sensor measurements embedding}
\label{subsec:voronoi}

Traditional regressive field reconstruction methods often represent sensor measurements as vectors \cite{Erichson2020, DUBOIS2022110733}, restricting the ability of machine learning models to generalize effectively to different sensor settings (e.g., varying sensor locations or layouts). In this work, we used the Voronoi tessellation to encode the sensor measurement information.

\begin{figure}[b]
  \centering
  \begin{minipage}[c]{0.67\textwidth}
    \includegraphics[width=\linewidth]{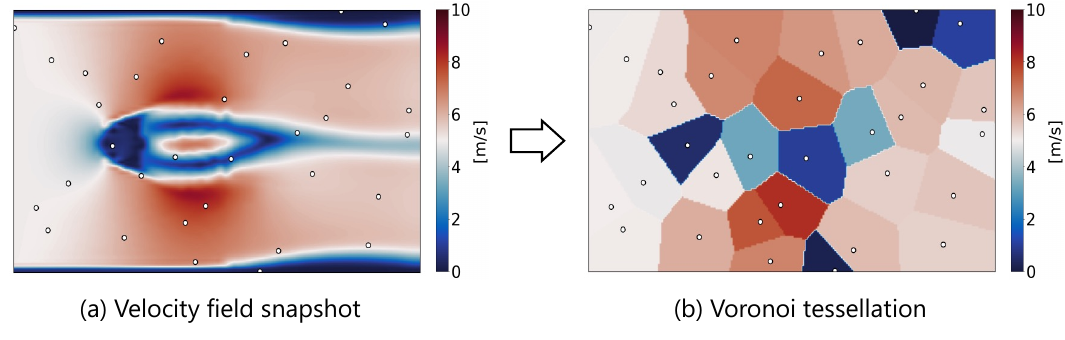}
  \end{minipage}\hfill
  \begin{minipage}[c]{0.3\textwidth}
    \caption{{Voronoi-Tessellation-Based Embedding:} (a) An original snapshot of the velocity field from the CFD dataset, with 32 randomly distributed sensors marked as white dots. (b) The corresponding Voronoi tessellation, computed from sensor locations and velocity values.}
    \label{fig:3}
  \end{minipage}
\end{figure}

Fig. \ref{fig:3} illustrates the Voronoi-tessellation-based embedding method. Given $p$ sensors located at coordinates $x_{c} \in \Omega$, the Voronoi tessellation divides $\Omega$ into $m$ Voronoi cells, denoted as $V = [V_1,V_2,...V_p]$. Each Voronoi cell $V_i$ is defined as:

\begin{equation}
\label{eq:voronoi}
V_i = \{ x \in \Omega \mid d(x, x_{c_i}) < d(x, x_{c_j}), \, \forall j \neq i \},
\end{equation}

where $d(\cdot, \cdot)$ denotes the Euclidean distance between two points. After the Voronoi tessellation process, all locations within each Voronoi cell are assigned the sensor value associated with that cell. This produces an embedding field with the same dimensions as the state field, defining the Voronoi embedding layer as $\Theta: \mathbb{R}^p \rightarrow \mathbb{R}^m$.

\subsection{Sinusoid embedding for temporal correlation}
\label{subsec:sinusoid}

The key feature processing component in trunk network $\mathcal{T}$ is the sinusoid embedding layer. This layer significantly enriches the feature space for the downstream MLP neural network. Particularly, for each time instance $t$ and $\tau_i$, the sinusoid embedding layer $\Phi: \mathbb{R} \rightarrow \mathbb{R}^d$ is defined as:

\begin{equation}
\label{eq:sinusoid}
\Phi(t) = \begin{bmatrix} 
\sin{(\omega_1 . t)} \\ 
\cos{(\omega_1 . t)} \\ 
\sin{(\omega_2 . t)} \\  
\cos{(\omega_2 . t)} \\ 
\vdots \\ 
\sin{(\omega_{d/2} . t)} \\
\cos{(\omega_{d/2} . t)} 
\end{bmatrix},
\end{equation}

where $\omega_k = \left(\frac{1}{10,000}\right)^{2k / d}$, and $d$ is the embedding dimension which is divisible by 2. In this particular problem, we choose $d = 64$, effectively transforming the 1-dimensional temporal domain into a 64-dimensional embedding space. Notably, $\Phi$ is highly efficient as it has no learnable parameters and Eq. \ref{eq:sinusoid} is inherently vectorizable for fast computation.

\subsection{Data and Training}
\label{sec:data}

\begin{figure}[t]
  \centering
  \begin{minipage}[c]{0.55\textwidth}
    \includegraphics[width=\linewidth]{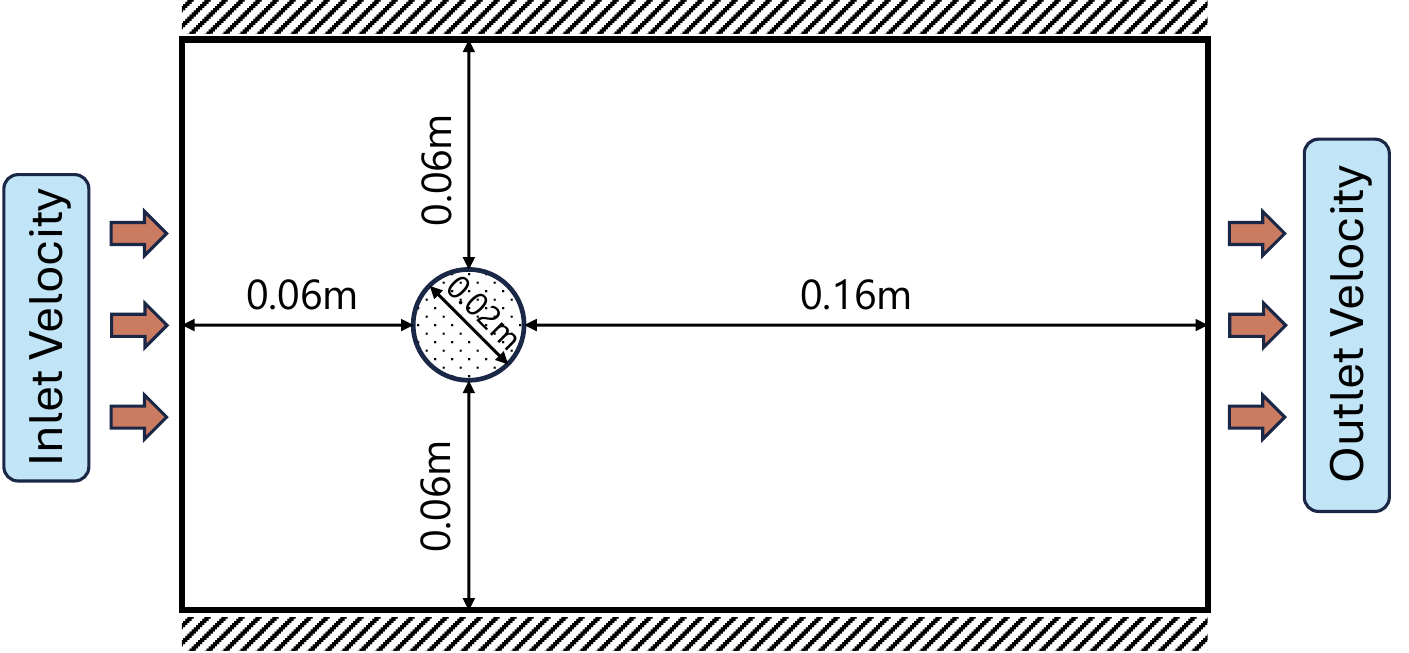}
  \end{minipage}\hfill
  \begin{minipage}[c]{0.4\textwidth}
    \caption{The geometry of the fluid flow around the circular cylinder problem as described in the CFD dataset.}
    \label{fig:4}
  \end{minipage}
\end{figure}

We trained FLRONet using the CFDBench cylinder dataset \cite{CFDBench}. The dataset contains numerical simulations of fluid flow around a circular cylinder (Fig. \ref{fig:4}). The examined domain has a physical size of $0.14\text{m} \times 0.24\text{m}$ and is tessellated in a grid of $140 \times 240$ pixels. There are 50 simulation cases in the dataset, each having a unique inlet velocity increasing incrementally from $0.1$ m/s to $5.0$ m/s. For each simulation case, the total duration is $1$ second, recorded over 1000 snapshots with a time interval of $0.001$ seconds between them. We randomly sample 32 locations across the domain to collect the velocity field, mimicking sparse sensor values that FLRONet uses as input. Fig. \ref{fig:4} depicts the geometry of the domain.

To construct the training data, we created observation windows containing five snapshots for the sensor data, with the time interval between two snapshots being 0.005 seconds, resulting in a window of 0.02 seconds. With this setup, 980 observation windows were generated for a single simulation case. Within each window, we randomly selected a single snapshot to record the complete velocity field, which served as the ground truth. To assess generalization, we randomly split 50 simulation cases into 45 cases for training and 5 cases for testing. The training set includes all inlet velocities except those reserved for testing, which are $3.5$m/s, $3.9$m/s, $4.2$m/s, $4.6$m/s, and $5.0$m/s. 

FLRONet was trained using the Adam optimizer with a learning rate of $10^{-3}$.

\subsection{Validation metrics}
\label{subsec:metrics}
In this work, we used the mean absolute error (MAE) to validate the accuracy of FLRONet. Given $\hat{\boldsymbol{u}}(t)$ as the reconstructed field at time $t$, $\mathbf{u}(t)$ is the corresponding ground truth. The MAE for a sample, defined by an observation window $\mathcal{W}$, is given as follows.

\begin{equation}
\label{eq:metrics} 
\text{MAE} = \frac{1}{|\mathcal{W}|}\sum_{t \in \mathcal{W}}{\|\hat{\boldsymbol{u}}(t) - \mathbf{u}(t)\|_1}, 
\end{equation}
where $|\mathcal{W}|$ is the number of available time steps in $\mathcal{W}$, and $\|\cdot\|_1$ is the $\ell_1$-norm. MAE is highly interpretable since it also has the unit of m/s.

\section{Results and discussion}
\label{sec:results}

We validate our FLRONet on the CFDBench dataset described in Section~\ref{sec:data}. For comparison, we benchmark FLRONet against the three-dimensional Fourier Neural Operator (FNO-3D). We also include two additional variants of FLRONet in which we replace FNO blocks by MLP blocks (FLRONet-MLP) and by UNet blocks (FLRONet-UNet). We conduct different experiments to assess FLRONet's ability in flow field reconstruction.

\subsection{Prediction accuracy}
\label{subsec:prediction_accuracy}

\begin{figure*}[tb!]
    \centering
    \includegraphics[width=0.95\textwidth]{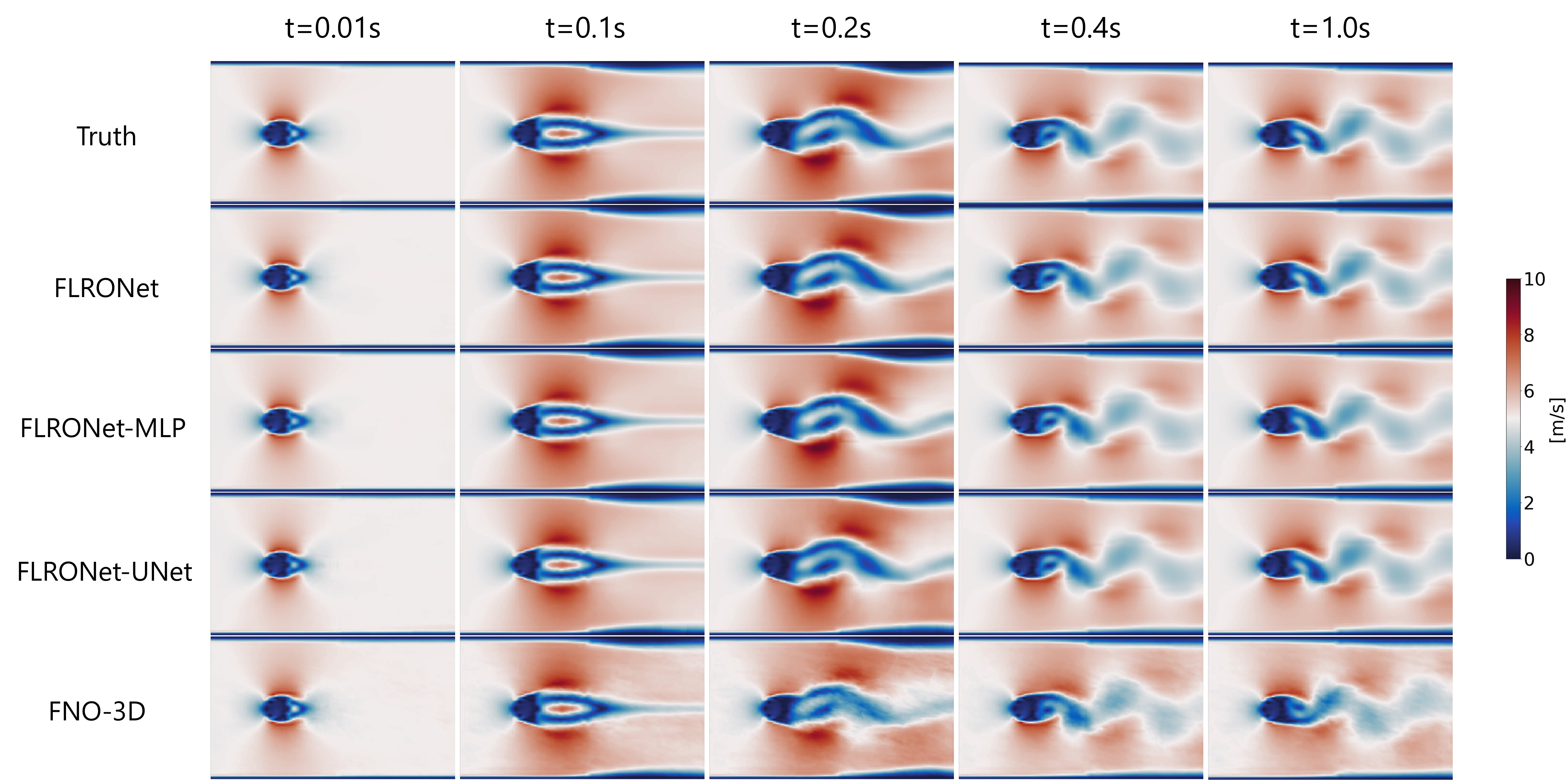}
    \caption{{Reconstructed fluid flow across architectures for the boundary condition \(v_0 = 3.5\) m/s.} The comparison highlights differences in reconstruction quality, with FLRONet showing the closest resemblance to the ground truth.}
    \label{fig:5}
\end{figure*}

\begin{figure*}[tb!]
    \centering
    \includegraphics[width=\textwidth]{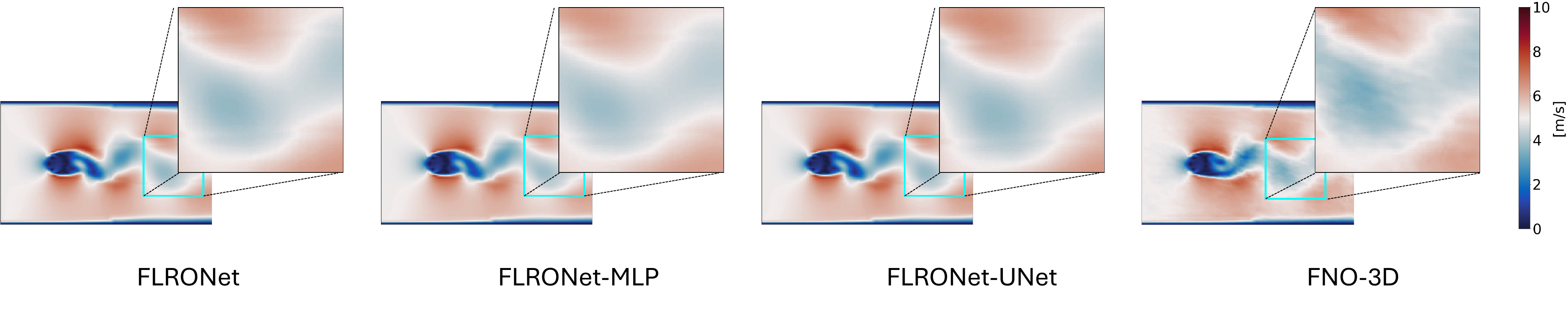}
    \caption{Detailed comparison of reconstructed fluid flow fields between FLRONet variants and FNO-3D for $v_0 = 3.5$ m/s at $t = 0.4$ s, focusing on the tail of the vortex-shedding region. The FNO-3D results exhibit noticeably fuzzier and noisier vortex structures, particularly at the interface regions, whereas FLRONet variants maintain sharper and more coherent features.}
    \label{fig:6}
\end{figure*}

In the first experiment, we validated FLRONet in a noise-free condition with no missing sensor measurements. Fig. \ref{fig:5} shows the reconstructed flow fields, ground truth, and corresponding MAE maps for FLRONet and other baselines on an unseen boundary condition $v_0 = 3.5$ m/s. Visually, all baselines capture the overall flow structure. However, while FLRONet architectures closely align with the ground truth from CFD, FNO-3D produces significantly noisier reconstructions, particularly at the tail of the vortex shedding feature (see Fig. \ref{fig:6}).

\begin{figure}[t]
  \centering
  \begin{minipage}[c]{0.45\textwidth}
    \includegraphics[width=\textwidth]{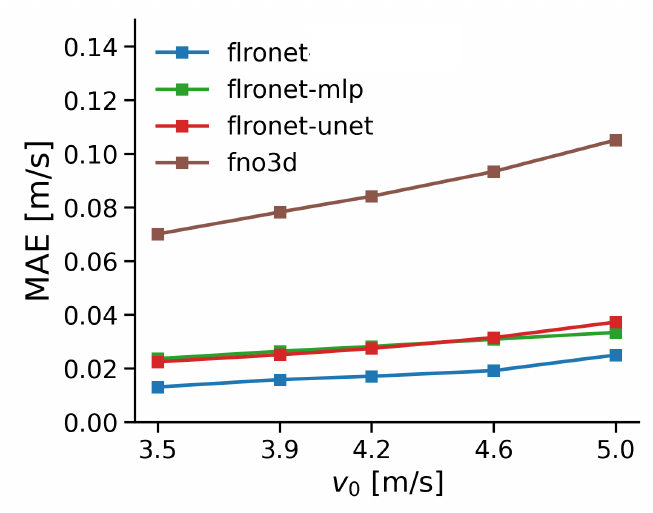}
  \end{minipage}\hfill
  \begin{minipage}[c]{0.5\textwidth}
    \caption{{Average reconstruction error across unseen inlet velocities.} FLRONet and its variants significantly outperform FNO-3D. FLRONet-MLP and FLRONet-UNet achieve similar accuracy. FLRONet surpasses all baselines with the lowest MAE values.}
    \label{fig:7}
  \end{minipage}
\end{figure}

\begin{figure}[t]
    \centering
    \begin{minipage}[c]{0.48\textwidth}
    \includegraphics[width=\textwidth]{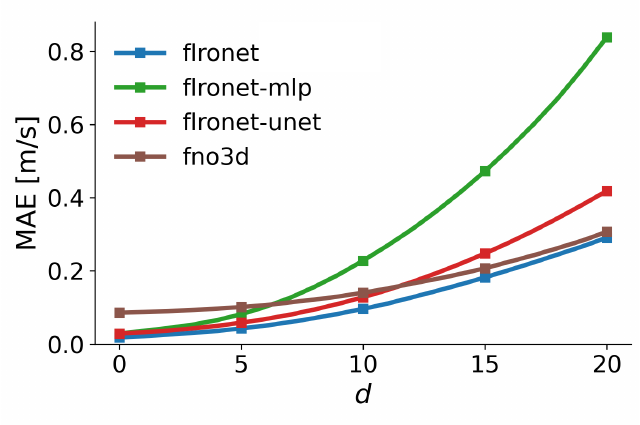}
    \end{minipage}\hfill
    \begin{minipage}[c]{0.5\textwidth}
    \caption{{Reconstruction performance under different numbers of lost sensors.} FLRONet is the most resilient to missing values as it leverages interpolation both in the Voronoi embedding layer and its operator-based design.}
    \label{fig:8}
    \end{minipage}
\end{figure}

In a more extensive analysis, we measured the average MAE of all architectures across all available time instances $t$. The evaluation was conducted on five test BCs that are never seen during training: $\{v_0 = 3.5, 3.9, 4.2, 4.6, 5.0\}$ m/s. As indicated in Fig. \ref{fig:7}, the fields reconstructed by FLRONet are consistently the most accurate, having the lowest reconstructed MAE at all values of $v_0$. Meanwhile, FNO-3D yields the highest errors. FLRONet-UNet and FLRONet-MLP exhibit similar performance, falling significantly behind FLRONet but still far better than FNO-3D. The MAE comparison reported in Tab. \ref{tab:model_comparison_mae} also confirms this observation.

An important observation across all architectures is that MAE values tend to increase as $v_0$ rises. Since higher $v_0$ results in greater velocity magnitudes of the entire system, the absolute error increases even for the same relative error. Furthermore, higher velocities correspond to greater Reynolds numbers. This implies more unstable and complex flow patterns, which in turn increases the reconstruction difficulty and error.

\subsection{Robustness to incomplete sensor measurements}
\label{subsec:missing_sensors}

In practice, it is not always possible to acquire all sensor data due to connection loss or sensor malfunction. In such scenarios, consistency in the reconstruction accuracy of the model is critical as it likely affects downstream tasks. For this reason, we are now validating FLRONet's performance in situations where sensor measurements are incomplete. Particularly, we randomly remove $d$ sensors and measure the accuracy of fields reconstructed from the remaining $(32 - d)$ sensors. Here, $d$ may go up to $20$.

Fig. \ref{fig:8} shows that FLRONet, FLRONet-UNet, and FNO-3D all maintain robust reconstruction accuracy. This robustness likely comes from the Voronoi embedding layer in these architectures. Leveraging sensor coordinates, the Voronoi embedding layer effectively encodes the spatial relationships, allows missing values to be easily interpolated back from their nearest neighbors. 
Conversely, FLRONet-MLP has no mechanisms to incorporate spatial information. When some sensors are lost, their values are unavoidably dropped to zeros, which severely distorts the reconstruction results. As a results, FLRONet-MLP experiences an exponential increase in MAE value with respect to $d$. 

Moreover, FNO-3D and FLRONet appear to exhibit slower accuracy degradation. This resilience stems from the fundamental paradigm of Fourier-based approaches, which treats input data as a \textit{continuous} function. Consequently, missing values are interpolated naturally through the function's continuity before being fed to the operator. In function space, the distance between the complete sensor input function \(\mathbf{y}(t)\) and the incomplete sensor input function \(\mathbf{y'}(t)\) remains small. A more detail quantitative report can be found in Tab. \ref{tab:model_comparison_mae}.

\begin{table*}[h]
\centering
\caption{Comparison of test-set mean absolute errors for FLRONet variants and FNO 3D}
\label{tab:model_comparison_mae}
\begin{tabular}{l c c c c c c c}
\toprule
Model & Accuracy 
      & \multicolumn{3}{c}{Noise robustness} 
      & \multicolumn{3}{c}{Dropout robustness} \\
\cmidrule(lr){3-5} \cmidrule(lr){6-8}
      & 
      & 5\%        & 10\%        & 20\% 
      & 5 sensors  & 10 sensors  & 20 sensors \\
\midrule
FLRONet–FNO
  & \textbf{0.01306}  
  & \textbf{0.05202} & 0.10491 & 0.20540 
  & \textbf{0.04271} & \textbf{0.09606} & \textbf{0.29092} \\
FLRONet–UNet
  & 0.02243  
  & 0.22391 & 0.33955 & 0.48029 
  & 0.05834 & 0.12793 & 0.41862 \\
FLRONet–MLP 
  & 0.02363  
  & 0.03440 & 0.06165 & \textbf{0.15942} 
  & 0.08149 & 0.22795 & 0.83885 \\
FNO 3D       
  & 0.07004  
  & 0.08060 & \textbf{0.10417} & 0.16696
  & 0.10124 & 0.14045 & 0.30688 \\
\bottomrule
\end{tabular}
\end{table*}

\subsection{Robustness to Imprecise Sensor Measurements}
\label{subsec:noise}

\begin{figure}[t!]
    \centering
    \begin{minipage}[c]{0.45\textwidth}
    \includegraphics[width=\textwidth]{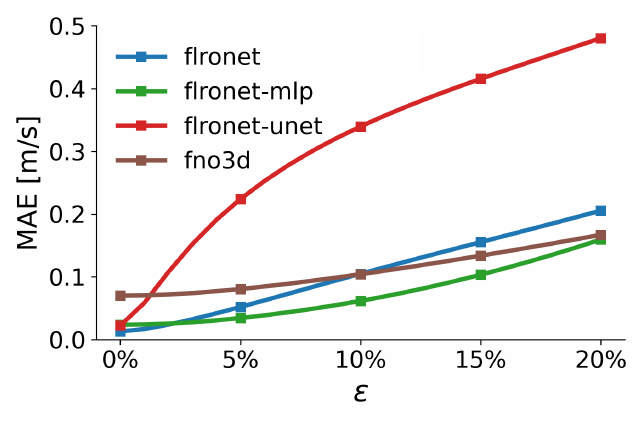}
    \end{minipage}\hfill
    \begin{minipage}[c]{0.5\textwidth}
    \caption{{Reconstruction performance under different noise levels.} FLRONet-UNet is highly sensitive to noise as it propagates noise directly to the final output through its skip connections. FLRONet withstands noise better than FNO-3D up to noise level $\epsilon = 10\%$.}
    \label{fig:9}
    \end{minipage}
\end{figure}

Beyond potential data loss, sensor measurements are often subjected to noise due to hardware limitations or environmental factors. In such cases, a noise-insensitive model is essential to ensure reliable predictions and prevent noise propagation to downstream tasks. In this experiment, we evaluate the resilience of FLRONet and other baselines to varying intensities of noise injected to sensor measurements. Specifically, Gaussian noise $\boldsymbol{\xi} \in \mathbb{R}^{p}$ is added to sensor observation $\mathbf{y} \in \mathbb{R}^{p}$:

\begin{equation}
\mathbf{y} \gets \mathbf{y} + \boldsymbol{\xi}, 
\quad 
\boldsymbol{\xi}_{i} \sim \mathcal{N}\left(0, (\epsilon |\mathbf{y}_{i}|)^2\right), \quad \forall i \in \{1, \dots, p\},
\label{eq:noise}
\end{equation}

\noindent where:
\renewcommand\labelitemi{}
\begin{itemize}
    \setlength{\itemsep}{0pt}
    \setlength{\parskip}{0pt}
    \item \(\mathbf{y}_{i}\): The velocity field measurement of the $i$-th sensor.
    \item \(\boldsymbol{\xi}_{i}\): Independent Gaussian noise applied to the sensor $i$, with mean $0$ and variance \((\epsilon |\mathbf{y}_{i}|)^2\).
    \item \(\epsilon \geq 0\): The relative noise level.
\end{itemize}

\vspace{5pt}
With the above formulation, we create a controlled environment to assess the resilience of each model to different levels of noise by simply adjusting $\epsilon$. Here, $\epsilon$ range from $0\%$ (noise-free) to up to $20\%$.

Although extremely sensitive to data loss as demonstrated in the previous section, Fig. \ref{fig:9} and Tab. \ref{tab:model_comparison_mae} indicate that FLRONet-MLP demonstrates strong resistance to noise at all levels. The lack of an interpolation mechanism due to the absence of a Voronoi embedding layer in FLRONet-MLP also means that noise affecting a sensor at a particular position does not propagate further in the spatial domain. The MLP branch network processes the sensor measurements independently and perturbations in a small subset of input features can be offset through the dense layers. Fig. \ref{fig:9} shows that while FLRONet-MLP is less competitive than FLRONet under noise-free inputs, it achieves the lowest MAE across all noise levels $\epsilon > 3\%$. Its error increases only gradually as $\epsilon$ increases from $0\%$ to $20\%$.

FLRONet and FNO-3D also effectively manage input noise thanks to their utilization of spectral transformations and mode truncations. In the frequency domain, both methods truncate redundant high-frequency modes, which are often dominated by noise. Theoretically, FNO-3D filters noise in both space and time while FLRONet only filters in space. However, there is no consistent difference in the performance of both methods in this practical application.

Unsurprisingly, FLRONet-UNet is the most noise-sensitive. Since skip connections transmit both noise and essential features directly to the output, measurement errors can easily bypass the necessary processes and degrade the reconstructed field. As a result, its MAE explodes even for small \(\epsilon\).

\subsection{Zero-shot super resolution in spatial domain}
\label{subsec:sr_space}

\begin{figure*}[t]
    \centering
    \includegraphics[width=0.87\textwidth]{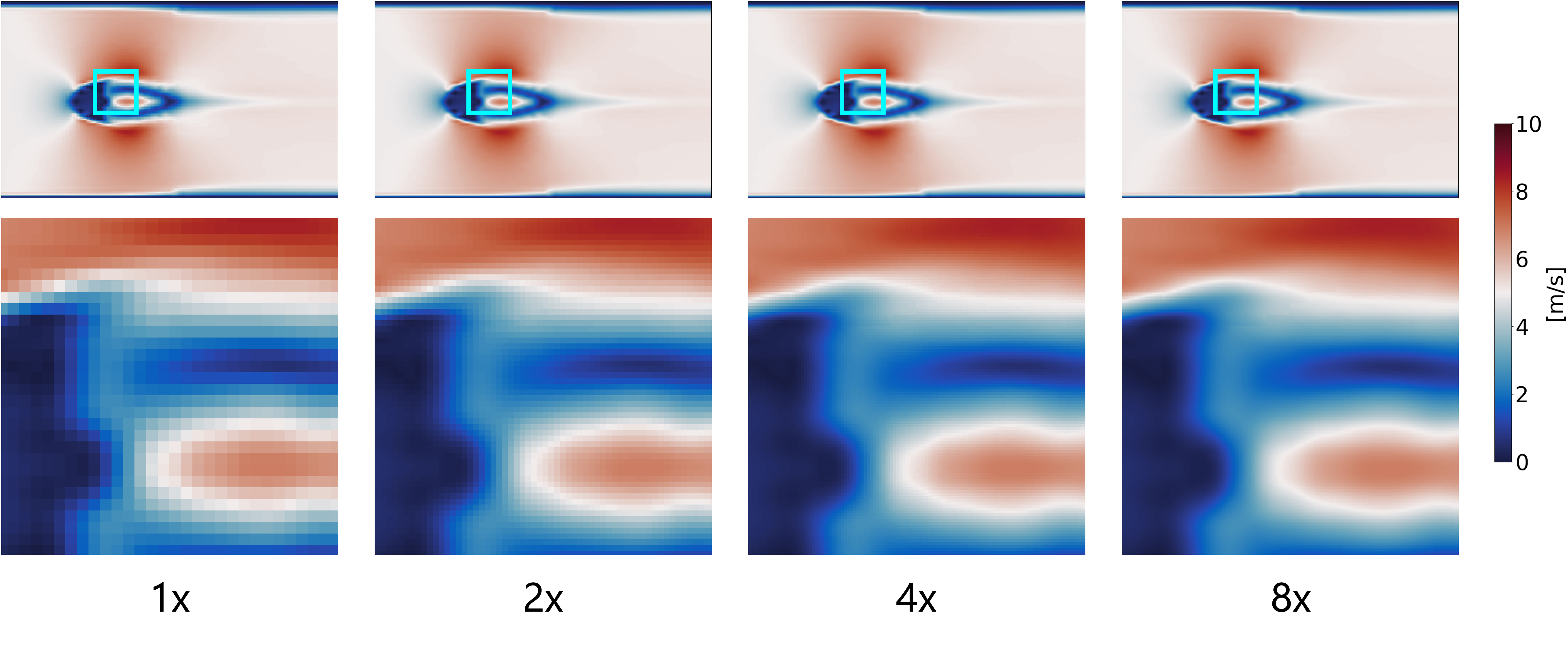}
    \caption{Super-resolution in space. The top row shows the full-field reconstruction, the bottom row provides magnified views of the highlighted regions to illustrate the improvement in spatial detail. We see the progressive reduction in pixelation as the resolution increases.
    }
    \label{fig:10}
\end{figure*}

\begin{figure*}[t]
    \centering
    \includegraphics[width=0.9\textwidth]{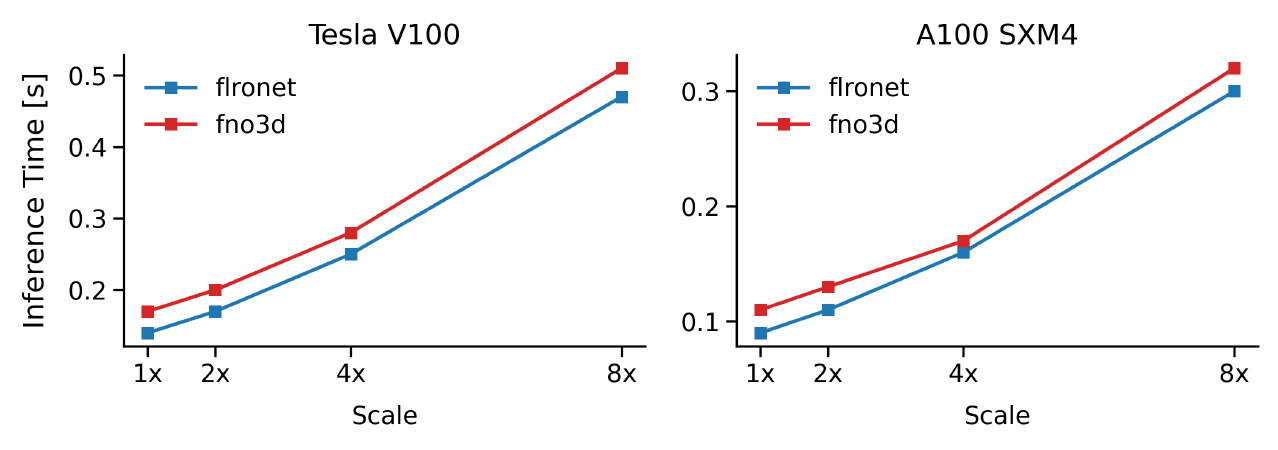}
    \caption{{Inference Time for Zero-Shot Super Resolution in Space.} The average inference time per time step for FLRONet and FNO-3D. FLRONet consistently achieves lower inference times than FNO-3D at all scales.}
    \label{fig:11}
\end{figure*}

In this section, we demonstrate the zero-shot super resolution capability of FLRONet in the spatial domain. In contrast to MLP-based or CNN-based architectures, which require retraining for different spatial scales, FLRONet is inherently resolution-independent. It can be trained on low-resolution data and infer on much higher resolution for free. This scalability is feasible because FLRONet reconstructs the physical field in the function space. During inference, it can naturally interpolate intermediate "pixels" using the predicted continuous function. As a result, knowledge learned from low-resolution data can be directly transferred to high-resolution settings.

Fig. \ref{fig:10} presents the upscaled reconstruction by FLRONet. Trained on the $(140 \times 240)$ resolution, FLRONet performs inference at $(280 \times 480)$, $(560 \times 960)$, and $(1120 \times 1920)$, corresponding to 2x, 4x, and 8x upscaling, respectively. The 8x reconstruction requires approximately 30GB of VRAM, which is well within the capacity of most consumer-grade GPUs. Higher-end GPUs, e.g., NVIDIA's A100 with 80GB VRAM or the H100 series, can accommodate 16x reconstructions, providing even more fine-grain details. FLRONet is able to reconstruct the complete vector field at arbitrary scales, constrained only by hardware memory.

The only method comparable to FLRONet in performing zero-shot super-resolution in space is FNO-3D. Both incur most of their computational cost from the (inverse) Fourier transforms. Fig. \ref{fig:11} presents the average inference time of each model on two popular mid-range GPUs: Tesla V100 (32GB) and A100 SMX4 (40GB). Inference time scales almost logarithmically. However, FLRONet empirically achieves lower inference time across all resolutions on both GPUs. The efficiency gap between the two models is more pronounced on Tesla V100. This result signifies the computational advantage of FLRONet, particularly on lower-FLOPS hardware.

\subsection{Zero-shot super resolution in temporal domain}
\label{subsec:sr_time}

\begin{figure*}[t]
    \centering
    \includegraphics[width=\textwidth]{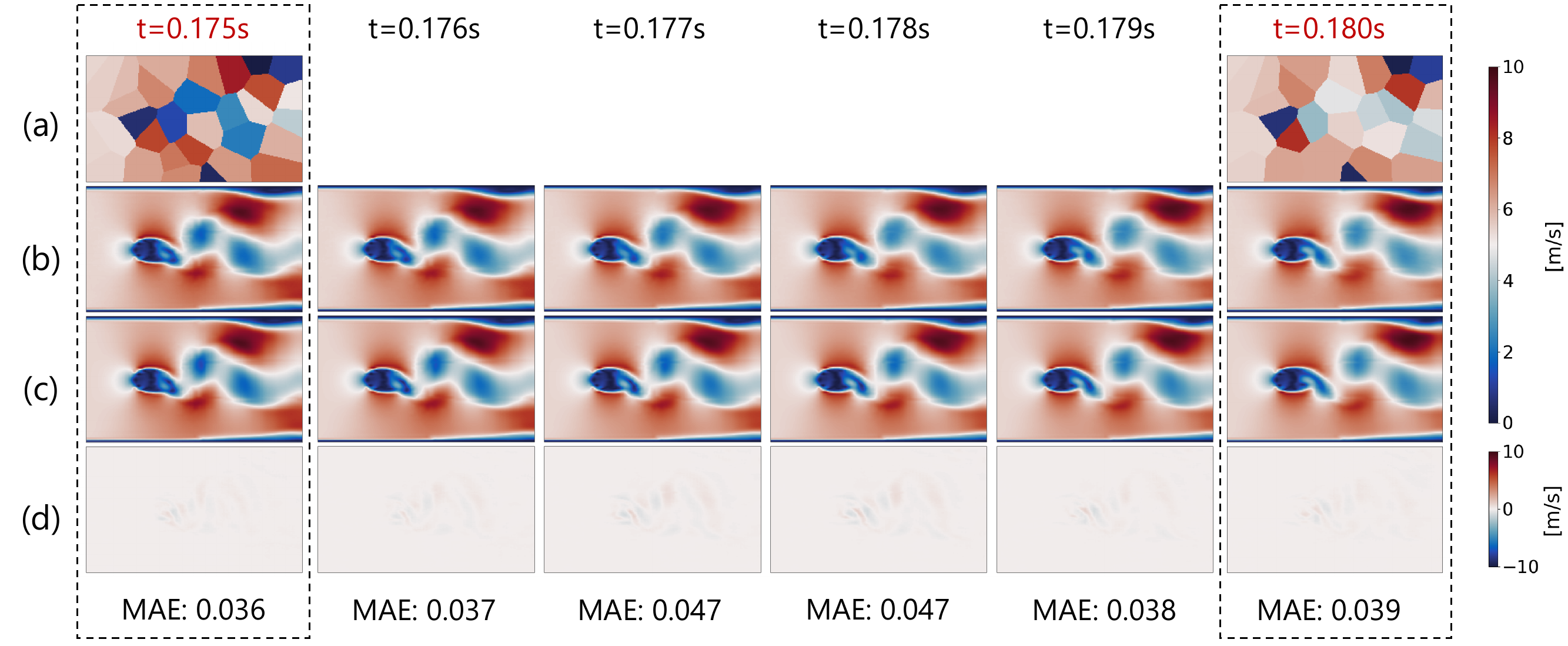}
    \caption{{Reconstruction results under the boundary condition $v_0 = 3.5$ m/s. (a)} Input Voronoi map observed at $t=0.175$s and $t=0.180$s. \textbf{(b)} Reconstructed velocity fields at all time instances. \textbf{(c)} Ground truth velocity fields at all time instances. \textbf{(d)} The error maps represent the absolute differences between rows (b) and (c). Average MAE values are provided below each time step to quantify the reconstruction accuracy.}
    \label{fig:12}
\end{figure*}

\begin{figure*}[t]
    \centering
    \includegraphics[width=0.6\textwidth]{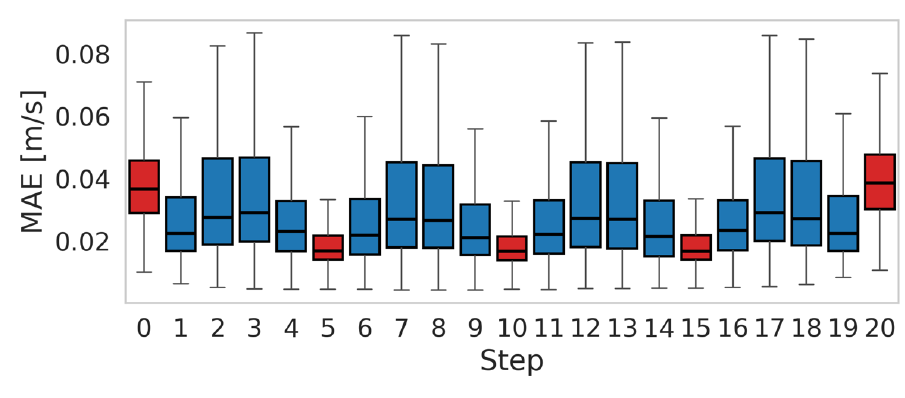}
    \caption{{Reconstruction performance at each step in the sliding window.} Steps corresponding to interior sensor inputs, i.e. indices $\{5; 10; 15\}$, consistently achieve the lowest errors, with minimal variability in MAE values. Predictions tend to be more accurate when made closer to the sensor inputs or towards the center of the sliding window. Predictions at indices $\{ 2; 3; 7; 8; 12; 13; 17; 18 \}$ have higher errors as they are temporally furthest from the sensor inputs at $\{ 0; 5; 10; 15; 20 \}$. This result aligns with the expected challenge of predicting further from observed data points. Notably, predictions at indices $\{ 0; 20 \}$ have higher errors despite being sensor input steps. This is because they are located at the edges of the sliding window and receive only partial temporal context.}
    \label{fig:13}
\end{figure*}

\begin{figure*}[tb!]
    \centering
    \includegraphics[width=\textwidth]{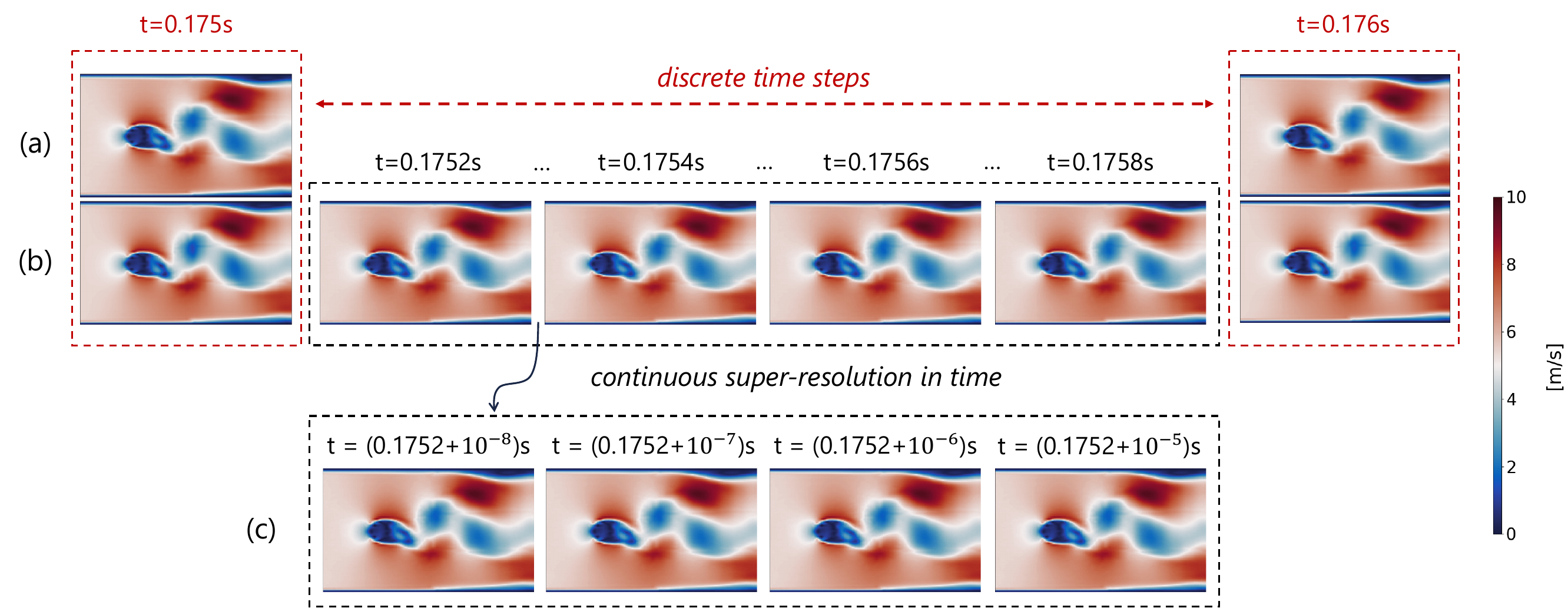}
    \caption{{Continuous super-resolution in time. (a)} The true velocity fields provided by the CFD dataset, which are only available at $t=0.175$s and $t=0.176$s. \textbf{(b)} The reconstructed velocity fields at intermediate time points: $t = 0.1750$s, $t = 0.1752$s, $t = 0.1754$s, $t = 0.1756$s, $t = 0.1758$s, and $t = 0.1760$s, highlighting the model's ability to continuously generate smooth transitions between consecutive discrete steps. \textbf{(c)} The capability of temporal super-resolution is further demonstrated by presenting reconstructions at finer time: $t = (0.1752 + 10^{-8})$s, $t = (0.1752 + 10^{-7})$s, $t = (0.1752 + 10^{-6})$s, and $t = (0.1752 + 10^{-5})$s.}
    \label{fig:14}
\end{figure*}

One of the most prominent features of FLRONet is its capability to perform zero-shot super-resolution in the temporal domain. This refers to the model's proficiency in reconstructing the flow field at any given time step within the observational window without the necessity of having the sensor observation at that specific time (see Fig. \ref{fig:12}). This capability is particularly useful in the case where vortex shedding occurs and the flow becomes highly unstable. In this scenario, it is required to capture finer temporal details, and this requirement can be easily fulfilled by FLRONet via its zero-shot temporal super-resolution.

In this experiment, FLRONet uses sensor data from five time steps, uniformly distributed within the temporal window $\mathcal{W}$, to reconstruct the flow field at any time $t \in \mathcal{W}$. Fig. \ref{fig:12} illustrates a particular result in which FLRONet uses sensor measurements at $\{\tau_1 = 0.165; \tau_2 = 0.170; \tau_3 = 0.175; \tau_4 = 0.180; \tau_5 = 0.185\}$ and reconstruct the flow field at $\{t = 0.175; t = 0.176; t = 0.177; t = 0.178; t = 0.179; t = 0.180\}$. Visually, the reconstructions appear identical to the ground truth from CFD. However, MAE increases as $t$ deviates from $\tau_i$ and reaches its lowest at $t = \tau_i$.

A more comprehensive experiment was conducted to evaluate the performance of FRLONet across all possible window $\mathcal{W}$. Particularly, the CFDBench dataset comprises $1,000$ time steps over a 1-second simulation. 
A $21$-step sliding window is applied sequentially across $1,000$ time steps with a stride of $1$, generating $980$ samples for each BC. Each sample consists of $21$ consecutive time steps indexed from $0$ to $20$. The input sensory frames are evenly distributed at indices $\{0; 5; 10; 15; 20 \}$. FLRONet aims to reconstruct all frames at indices $0 \rightarrow 20$ and the results are compared against the CFD ground truth. 

Fig. \ref{fig:13} illustrates how reconstruction accuracy varies with the relative position of the target frame. Interior sensor time steps, i.e. at indices $\{5; 10; 15\}$, consistently yield lower errors, while frames distant to sensor observations exhibit higher errors. Edge frames, i.e. at indices $\{0; 20\}$, usually have higher errors despite being sensor frames likely because they only have access to half of the temporal context. These results well align with our intuition. Although there are some minor differences in accuracy depending on the relative temporal position of the target frame, test MAE values rarely exceed $0.08$ m/s. Considering the velocity that may go up to $10$ m/s, this error is only about $0.8\%$. 

Despite being trained on discrete time intervals $\Delta_t = 10^{-3}$s, FLRONet can make reconstruction at any continuous value $t \in \mathcal{W}$. This property enables reconstruction at arbitrary continuous time points within the observation window, with a theoretical precision limited only by the data’s floating-point representation. Row (b) in Fig. \ref{fig:14} illustrates the FLRONet-reconstructed fields at multiple intermediate time steps with $\Delta_t = 10^{-4}$. Row (c) even goes further by delivering reconstruction with $\Delta_t = 10^{-5}$s, $\Delta_t = 10^{-6}$s, $\Delta_t = 10^{-7}$s, and $\Delta_t = 10^{-8}$s.  Although the ground truth at such fine resolution is unavailable for a quantitative validation, the reconstructed fields appear visually valid, showing no signs of numerical instability or artifacts. 

\subsection{Discussion and implications}
\label{subsec:discussion}

The validation study from Section \ref{subsec:prediction_accuracy} to Section \ref{subsec:sr_time} confirms the high reconstruction quality of FLRONet. Here, we want to emphasize the efficient zero-shot super-resolution capability in both space and time of FLRONet and its implications. Real-time high-frequency data acquisition is often impractical in industrial settings. FLRONet allows infinitely fine-scale modeling of mechanical systems from extremely sparse sensor measurements in both space and time. In our experiments, FLRONet only observes $32 \times 5 = 160$ data points per window $\mathcal{W}$ yet computes up to $1120 \times 1920 = 2{,}150{,}400$ data points at any arbitrary frame at $t \in \mathcal{W}$ using just a mid-range hardware. Moreover, its mesh-invariance property also reduces the training costs by allowing the model to be trained on inexpensive, low-resolution data yet to be able to deliver infinitely high-fidelity predictions.

The reconstruction error behavior of FLRONet, illustrated in Fig.~\ref{fig:13}, also carries important implications. For high-accuracy reconstruction, the target time stamps should be positioned as close as possible to the available sensor observations. This suggests that a denser temporal sampling becomes necessary when very fine-grained accuracy is required. However, this introduces a fundamental trade-off between reconstruction fidelity and the cost of collecting additional sensor data. An interesting direction for future work is to investigate optimal sampling strategies, whether uniform or nonuniform, that can maximize reconstruction accuracy while minimizing data acquisition costs.

Finally, it is important to note that all reconstructions in this study are performed strictly within the observation window $[\tau_0, \tau_n]$, corresponding to a temporal interpolation task. Temporal extrapolation, i.e., forecasting beyond the last observation time ($t > \tau_n$), is not addressed here. Such an extrapolation poses additional challenges related to error accumulation, stability, and uncertainty growth and will be the subject of future work.

\section{Conclusion}
\label{sec:conclusion}
In this paper, we propose FLRONet, a deep operator learning architecture designed for reconstructing high-fidelity fluid flow fields from sparse sensor measurements. Our findings indicate that FLRONet produces field reconstructions closely aligned with CFD simulations while demonstrating high robustness against noisy and incomplete sensor data. Notably, FLRONet achieves super-resolution in both space and time without retraining, an ability that is not available in conventional deep learning-based field reconstruction techniques that rely on MLPs or CNNs. Our future work on FLRONet will focus on evaluating its temporal extrapolation capability, specifically assessing how well the model can forecast flow fields at time points beyond the observation window. Additionally, studying its generalizability to real-world data remains an open research direction.

\bibliographystyle{unsrt}  
\bibliography{main}  

@article{li2024deep,
title = {Deep learning reconstruction of high-Reynolds-number turbulent flow field around a cylinder based on limited sensors},
journal = {Ocean Engineering},
volume = {304},
pages = {117857},
year = {2024},
issn = {0029-8018},
doi = {https://doi.org/10.1016/j.oceaneng.2024.117857},
author = {Rui Li and Baiyang Song and Yaoran Chen and Xiaowei Jin and Dai Zhou and Zhaolong Han and Wen-Li Chen and Yong Cao},
}

@article{Erichson2020,
author = {Erichson, N. Benjamin  and Mathelin, Lionel  and Yao, Zhewei  and Brunton, Steven L.  and Mahoney, Michael W.  and Kutz, J. Nathan },
title = {Shallow neural networks for fluid flow reconstruction with limited sensors},
journal = {Proceedings of the Royal Society A: Mathematical, Physical and Engineering Sciences},
volume = {476},
number = {2238},
pages = {20200097},
year = {2020},
doi = {https://doi.org/10.1098/rspa.2020.0097},
}

@inproceedings{
        luo2024continuous,
        title={Continuous Field Reconstruction from Sparse Observations with Implicit Neural Networks},
        author={Xihaier Luo and Wei Xu and Balu Nadiga and Yihui Ren and Shinjae Yoo},
        booktitle={The Twelfth International Conference on Learning Representations},
        year={2024},
        }

@article{Wen2021UFNOA,
title = {U-FNO—An enhanced Fourier neural operator-based deep-learning model for multiphase flow},
journal = {Advances in Water Resources},
volume = {163},
pages = {104180},
year = {2022},
issn = {0309-1708},
doi = {https://doi.org/10.1016/j.advwatres.2022.104180},
url = {https://www.sciencedirect.com/science/article/pii/S0309170822000562},
author = {Gege Wen and Zongyi Li and Kamyar Azizzadenesheli and Anima Anandkumar and Sally M. Benson},
}

@inproceedings{Li2020MultipoleGN,
  abbr = {NeurIPS},
  bibtex_show = {true},
  author = {Li, Zongyi and Kovachki, Nikola B and Azizzadenesheli, Kamyar and Liu, Burigede and Stuart, Andrew M and Bhattacharya, Kaushik and Anandkumar, Anima},
  booktitle = {Advances in Neural Information Processing Systems (NeurIPS)},
  title = {Multipole Graph Neural Operator for Parametric Partial Differential Equations},
  html = {https://papers.nips.cc/paper/2020/hash/4b21cf96d4cf612f239a6c322b10c8fe-Abstract.html},
  volume = {33},
  year = {2020}
}

@article{Li2021PhysicsInformedNO,
author = {Li, Zongyi and Zheng, Hongkai and Kovachki, Nikola and Jin, David and Chen, Haoxuan and Liu, Burigede and Azizzadenesheli, Kamyar and Anandkumar, Anima},
title = {Physics-Informed Neural Operator for Learning Partial Differential Equations},
year = {2024},
issue_date = {September 2024},
publisher = {Association for Computing Machinery},
address = {New York, NY, USA},
volume = {1},
number = {3},
url = {https://doi.org/10.1145/3648506},
doi = {10.1145/3648506},

journal = {ACM / IMS J. Data Sci.},
month = may,
articleno = {9},
numpages = {27}
}

@article{PENG2023108539,
title = {A hybrid deep learning framework for unsteady periodic flow field reconstruction based on frequency and residual learning},
journal = {Aerospace Science and Technology},
volume = {141},
pages = {108539},
year = {2023},
issn = {1270-9638},
doi = {https://doi.org/10.1016/j.ast.2023.108539},
url = {https://www.sciencedirect.com/science/article/pii/S1270963823004364},
author = {Xingwen Peng and Xingchen Li and Xiaoqian Chen and Xianqi Chen and Wen Yao},
}

@article{LIU2021120684,
title = {Supervised learning method for the physical field reconstruction in a nanofluid heat transfer problem},
journal = {International Journal of Heat and Mass Transfer},
volume = {165},
pages = {120684},
year = {2021},
issn = {0017-9310},
doi = {https://doi.org/10.1016/j.ijheatmasstransfer.2020.120684},
url = {https://www.sciencedirect.com/science/article/pii/S0017931020336206},
author = {Tianyuan Liu and Yunzhu Li and Qi Jing and Yonghui Xie and Di Zhang},
}

@article{ZHAO2024108619,
title = {RecFNO: A resolution-invariant flow and heat field reconstruction method from sparse observations via Fourier neural operator},
journal = {International Journal of Thermal Sciences},
volume = {195},
pages = {108619},
year = {2024},
issn = {1290-0729},
doi = {https://doi.org/10.1016/j.ijthermalsci.2023.108619},
url = {https://www.sciencedirect.com/science/article/pii/S1290072923004805},
author = {Xiaoyu Zhao and Xiaoqian Chen and Zhiqiang Gong and Weien Zhou and Wen Yao and Yunyang Zhang},
}

@article{DUBOIS2022110733,
title = {Machine learning for fluid flow reconstruction from limited measurements},
journal = {Journal of Computational Physics},
volume = {448},
pages = {110733},
year = {2022},
issn = {0021-9991},
doi = {https://doi.org/10.1016/j.jcp.2021.110733},
url = {https://www.sciencedirect.com/science/article/pii/S0021999121006288},
author = {Pierre Dubois and Thomas Gomez and Laurent Planckaert and Laurent Perret},
}

@article{Nguyen_review,
    author = {Nguyen, Phong C. H. and Choi, Joseph B. and Udaykumar, H. S. and Baek, Stephen},
    title = {Challenges and Opportunities for Machine Learning in Multiscale Computational Modeling},
    journal = {Journal of Computing and Information Science in Engineering},
    volume = {23},
    number = {6},
    pages = {060808},
    year = {2023},
    month = {05},
    issn = {1530-9827},
    doi = {10.1115/1.4062495},
    url = {https://doi.org/10.1115/1.4062495},
    eprint = {https://asmedigitalcollection.asme.org/computingengineering/article-pdf/23/6/060808/7014978/jcise\_23\_6\_060808.pdf},
}

@article{MONS2016255,
title = {Reconstruction of unsteady viscous flows using data assimilation schemes},
journal = {Journal of Computational Physics},
volume = {316},
pages = {255-280},
year = {2016},
issn = {0021-9991},
doi = {https://doi.org/10.1016/j.jcp.2016.04.022},
url = {https://www.sciencedirect.com/science/article/pii/S0021999116300638},
author = {V. Mons and J.-C. Chassaing and T. Gomez and P. Sagaut},
}

@article{Loiseau_Noack_Brunton_2018, title={Sparse reduced-order modelling: sensor-based dynamics to full-state estimation}, volume={844}, DOI={10.1017/jfm.2018.147}, journal={Journal of Fluid Mechanics}, author={Loiseau, Jean-Christophe and Noack, Bernd R. and Brunton, Steven L.}, year={2018}, pages={459–490}}

@article{CHENG2024112581,
title = {Efficient deep data assimilation with sparse observations and time-varying sensors},
journal = {Journal of Computational Physics},
volume = {496},
pages = {112581},
year = {2024},
issn = {0021-9991},
doi = {https://doi.org/10.1016/j.jcp.2023.112581},
url = {https://www.sciencedirect.com/science/article/pii/S0021999123006769},
author = {Sibo Cheng and Che Liu and Yike Guo and Rossella Arcucci},
}

@article{JMLR:v24:21-1524,
  author  = {Nikola Kovachki and Zongyi Li and Burigede Liu and Kamyar Azizzadenesheli and Kaushik Bhattacharya and Andrew Stuart and Anima Anandkumar},
  title   = {Neural Operator: Learning Maps Between Function Spaces With Applications to PDEs},
  journal = {Journal of Machine Learning Research},
  year    = {2023},
  volume  = {24},
  number  = {89},
  pages   = {1--97},
  url     = {http://jmlr.org/papers/v24/21-1524.html}
}

@inproceedings{DBLP:journals/corr/abs-2010-08895,
  abbr = {ICLR},
  bibtex_show = {true},
  author = {Li, Zongyi and Kovachki, Nikola B and Azizzadenesheli, Kamyar and Liu, Burigede and Bhattacharya, Kaushik and Stuart, Andrew M and Anandkumar, Anima},
  title = {Fourier Neural Operator for Parametric Partial Differential Equations},
  booktitle = {International Conference on Learning Representations (ICLR)},
  volume = {9},
  publisher = {OpenReview.net},
  year = {2021},
  html = {https://iclr.cc/virtual/2021/poster/3281},
  selected = {true}
}

@article{lu2021learning,
  title   = {Learning nonlinear operators via {DeepONet} based on the universal approximation theorem of operators},
  author  = {Lu, Lu and Jin, Pengzhan and Pang, Guofei and Zhang, Zhongqiang and Karniadakis, George Em},
  journal = {Nature Machine Intelligence},
  volume  = {3},
  number  = {3},
  pages   = {218--229},
  year    = {2021}
}

@article{Ronneberger2015,
   author = {Olaf Ronneberger and Philipp Fischer and Thomas Brox},
   city = {Cham},
   editor = {Nassir Navab and Joachim Hornegger and William M Wells and Alejandro F Frangi},
   isbn = {978-3-319-24574-4},
   journal = {Medical Image Computing and Computer-Assisted Intervention – MICCAI 2015},
   pages = {234-241},
   publisher = {Springer International Publishing},
   title = {U-Net: {C}onvolutional Networks for Biomedical Image Segmentation},
   year = {2015},
}

@online{CFDBench,
  title   = {CFDBench: A Large-Scale Benchmark for Machine Learning Methods in Fluid Dynamics},
  author  = {Luo, Yining and Chen, Yingfa and Zhang, Zhen},
  year    = {2023},
  url     = {https://github.com/luo-yining/CFDBench},
  note    = {Accessed: 2025-10-29}
}

@article{Zhu2025,
    author = {Zhu, Tong and Liu, Dehao and Lu, Yanglong},
    title = {Finite-Volume Physics-Informed U-Net for Flow Field Reconstruction With Sparse Data},
    journal = {Journal of Computing and Information Science in Engineering},
    volume = {25},
    number = {7},
    pages = {071004},
    year = {2025},
    month = {04},
    issn = {1530-9827},
    doi = {10.1115/1.4067583},
    url = {https://doi.org/10.1115/1.4067583},
    eprint = {https://asmedigitalcollection.asme.org/computingengineering/article-pdf/25/7/071004/7422303/jcise-24-1431.pdf},
}

@inproceedings{Yao2023,
  author={Weijie Yao and Wei Peng and Xiaoya Zhang and Wen Yao},
  title={A Method for Flow Field Reconstruction based on Fourier Neural Operator Network},
  year={2023},
  cdate={1672531200000},
  pages={408-412},
  url={https://doi.org/10.1145/3594315.3594349},
  booktitle={ICCAI},
}

@article{Moya2023,
    author = {Moya, Christian and Lin, Guang},
    title = {Bayesian, Multifidelity Operator Learning for Complex Engineering Systems–A Position Paper},
    journal = {Journal of Computing and Information Science in Engineering},
    volume = {23},
    number = {6},
    pages = {060814},
    year = {2023},
    month = {06},
    issn = {1530-9827},
    doi = {10.1115/1.4062635},
    url = {https://doi.org/10.1115/1.4062635},
    eprint = {https://asmedigitalcollection.asme.org/computingengineering/article-pdf/23/6/060814/7018943/jcise\_23\_6\_060814.pdf},
}

@article{lin2021operator,
  title={Operator learning for predicting multiscale bubble growth dynamics},
  author={Lin, Chensen and Li, Zhen and Lu, Lu and Cai, Shengze and Maxey, Martin and Karniadakis, George Em},
  journal={The Journal of Chemical Physics},
  volume={154},
  number={10},
  year={2021},
  publisher={AIP Publishing}
}

@article{goswami2022physics,
  title={A physics-informed variational DeepONet for predicting crack path in quasi-brittle materials},
  author={Goswami, Somdatta and Yin, Minglang and Yu, Yue and Karniadakis, George Em},
  journal={Computer Methods in Applied Mechanics and Engineering},
  volume={391},
  pages={114587},
  year={2022},
  publisher={Elsevier}
}

@article{Liu2025,
    author = {Liu, Daxin and Guo, Xuxin and Liu, Zhenyu and Tan, Jianrong},
    title = {An Image Generator Enhanced Deep Operator Network for Predicting the Geometry Deformations in Contact Problems with Random Rough Surfaces},
    journal = {Journal of Computing and Information Science in Engineering},
    pages = {1-26},
    year = {2025},
    month = {04},
    issn = {1530-9827},
    doi = {10.1115/1.4068456},
    url = {https://doi.org/10.1115/1.4068456},
    eprint = {https://asmedigitalcollection.asme.org/computingengineering/article-pdf/doi/10.1115/1.4068456/7466503/jcise-24-1448.pdf},
}

@article{Faroughi2024,
    author = {Faroughi, Salah A. and Pawar, Nikhil M. and Fernandes, Célio and Raissi, Maziar and Das, Subasish and Kalantari, Nima K. and Kourosh Mahjour, Seyed},
    title = {Physics-Guided, Physics-Informed, and Physics-Encoded Neural Networks and Operators in Scientific Computing: Fluid and Solid Mechanics},
    journal = {Journal of Computing and Information Science in Engineering},
    volume = {24},
    number = {4},
    pages = {040802},
    year = {2024},
    month = {01},
    issn = {1530-9827},
    doi = {10.1115/1.4064449},
    url = {https://doi.org/10.1115/1.4064449},
    eprint = {https://asmedigitalcollection.asme.org/computingengineering/article-pdf/24/4/040802/7383247/jcise\_24\_4\_040802.pdf},
}

@article{Fukami,
author = {Fukami, Kai and An, Byungjin and Nohmi, Motohiko and Obuchi, Masashi and Taira, Kunihiko},
year = {2022},
month = {08},
pages = {},
title = {Machine-Learning-Based Reconstruction of Turbulent Vortices From Sparse Pressure Sensors in a Pump Sump},
volume = {144},
journal = {Journal of Fluids Engineering},
doi = {10.1115/1.4055178}
}

\end{document}